# PENERAPAN TEKNIK ANALISIS CITRA UNTUK IDENTIFIKASI TEKSTIL

Andrian Wijayono[1] & Valentinus Galih Vidia Putra[1]

Textile Engineering Departement, Politeknik STTT Bandung, Indonesia[1]

**Abstrak:** Teknik analisis citra komputer yang digunakan untuk mengidentifikasi produk tekstil, disajikan dalam bab ini, bersamaan dengan tinjauan singkat perkembangan historis metode tersebut. Metode koreksi gambar otomatis dan semi otomatis dijelaskan, yang sering digunakan untuk identifikasi produk tekstil benang, dan juga dapat digunakan untuk mengidentifikasi sambungan benang yang disambung.

**Kata Kunci:** *digital image acquirement, image modelling, image quality improvement, image correction methods, median filtration, Laplace filter, threshold function, autocorrelation, Fourier transform, erosion, dilatation, digitalisation algorithm.*

**1. PENDAHULUAN**

Perkembangan teknik komputer yang dinamis menciptakan kemungkinan besar untuk penerapannya, termasuk mengidentifikasi dan mengukur dimensi geometris benda-benda yang sangat kecil termasuk benda tekstil. Menggunakan analisis citra digital memungkinkan analisis parameter struktural dasar produk tekstil benang yang lebih rinci seperti *thickness*, *hairiness* dan jumlah *twist*. Terlebih lagi, teknik ini juga memungkinkan dilakukannya pengukuran sifat-sifat lain dari struktur eksternal produk tekstil benang, seperti parameter *twist* dan nomor benang [6]. Proses identifikasi parameter benang struktural merupakan masalah yang signifikan, mengingat banyak dilakukannya penelitian ilmiah dan praktik industri saat ini. Berdasarkan literatur yang mempertimbangkan masalah ini, kita dapat menyatakan bahwa teknik pemrosesan gambar memungkinkan pengambilan penampang serat longitudinal & transversal yang akan diperoleh, diameter serat yang akan dinilai lebih lanjut, dan gambar produk tekstil benang memungkinkan pengamatan terhadap kemungkinan cacat benang dan penentuan penyebabnya. Gambar dua dimensi dan tiga dimensi yang diperoleh dapat membantu menciptakan produk tekstil, termasuk gambar sambungan benang dan perkiraan ketepatannya. Berdasarkan literatur, dapat disimpulkan bahwa pengolahan gambar digital dari citra produk tekstil umumnya menggunakan pemrosesan komputer gambar 2D.

## 2. KONSEP DASAR PENGUNAAN PENGOLAHAN CITRA

Analisis citra digital dua dimensi didasarkan pada pemrosesan perolehan citra dengan penggunaan komputer. Citra digital digambarkan oleh matriks dua dimensi, serta disajikan dalam bentuk bilangan real atau imajiner dengan jumlah byte yang pasti [1]. Sistem pengolahan citra digital dapat disajikan secara skematik seperti yang ditunjukkan pada Gambar-1.

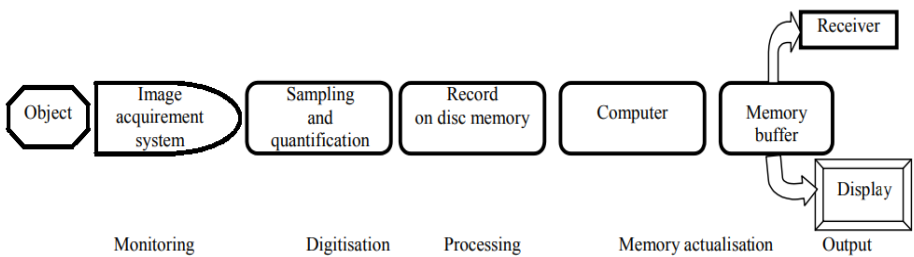

Gambar-1 Skema dari urutan fungsi dasar yang direalisasikan oleh suatu sistem pengolahan citra digital [1].

Pengolahan citra digital meliputi:
- perolehan dan pemodelan citra,

- Peningkatan kualitas gambar dan penyorotan fitur pembedanya,
- mengembalikan fitur gambar yang dikehendaki, dan
- kompresi data citra.

Pemodelan gambar [1] didasarkan pada digitalisasi citra sebenarnya. Proses ini terdiri dari sampling dan kuantifikasi gambar. Citra digital dapat digambarkan dalam bentuk matriks dua dimensi, yang unsur-unsurnya mencakup nilai kuantitatif dari fungsi intensitas, yang disebut tingkat abu-abu. Citra digital ditentukan oleh resolusi gambar spasial dan resolusi tingkat abu-abu. Unsur terkecil dari citra digital disebut pixel. Jumlah piksel dan jumlah tingkat kecerahan mungkin tidak terbatas, walaupun saat menyajikan data teknik komputer, biasanya menggunakan nilai yang merupakan perkalian dari angka 2, misalnya 512 × 512 piksel dan 256 tingkat abu-abu.

Peningkatan kualitas gambar dan penyorotan fitur utamanya adalah teknik aplikasi yang paling sering digunakan untuk pemrosesan gambar. Proses peningkatan kualitas gambar tidak meningkatkan informasi penting yang ditunjukkan oleh data gambar, namun meningkatkan rentang dinamis fitur yang dipilih dari objek yang diperoleh, yang memudahkan pendeteksiannya. Berikut ini adalah operasi dilakukan selama peningkatan kualitas gambar:

- Perubahan sistem *grey level* dan perbaikan *contrast*,
- *edge exposition*,
- *pseudo-colorisation*,
- peningkatan ketajaman gambar,
- menurunkan *noise level*,
- *space filtration*,
- *interpolation* dan *magnification*, dan
- kompensasi pengaruh faktor gangguan lainnya, misalnya kemungkinan *under-exposure*.

## 3. METODE KOREKSI CITRA

Metode histogram adalah salah satu prosedur koreksi gambar yang paling sederhana. Woźnicki [1] mendefinisikan histogram sebagai distribusi statistik dari fitur *grey level* tertentu yang terjadi pada citra digital. Prosedur ini digunakan terutama untuk meningkatkan *contrast*, meningkatkan bayangan gambar yang terlalu terang (*over exposure*), dan menyoroti gambar yang terlalu gelap (*under-exposure*) [1]. Modifikasi histogram akan mengubah fungsi edge histogram. Contoh penerapan metode ini untuk meratakan histogram ditunjukkan pada Gambar-2 dan Gambar-3. Metode rata-rata histogram sering kali digunakan dengan menggunakan intuisi.

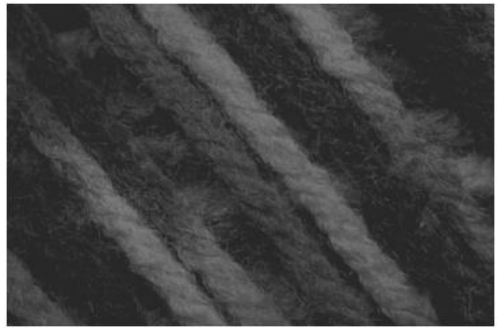

Gambar-2 Citra kain tenun sebelum dilakukan modifikasi histogram.

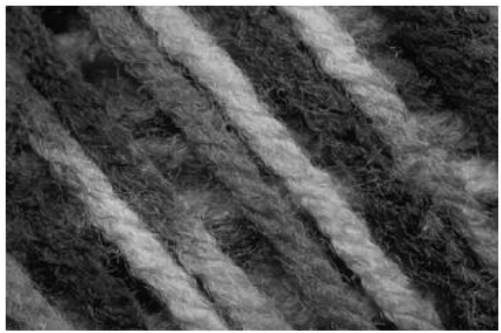

Gambar-3 Citra kain tenun dari Gambar-2 yang telah dilakukan operasi modifikasi histogram.

Metode lain koreksi gambar adalah dengan menggunakan fungsi rata-rata *brightness*, yang disebut prosedur *averaging mask*. Penerapan prosedur ini akan mengganti *brgihtness* asli pada piksel menjadi *brightness* rata-rata dari hasil teknik ini. Prosedur ini bertujuan untuk menghilangkan deformasi kecil pada citra digital yang dimanifestasikan saat adanya titik *exposure* atau cacat titik [1]. Penggunaan prosedur *averaging mask* dalam pemrosesan citra digital meningkatkan ketajaman bentuk objek. Hasil penggunaan *mask* tersebut seperti ditunjukkan pada Gambar-4.

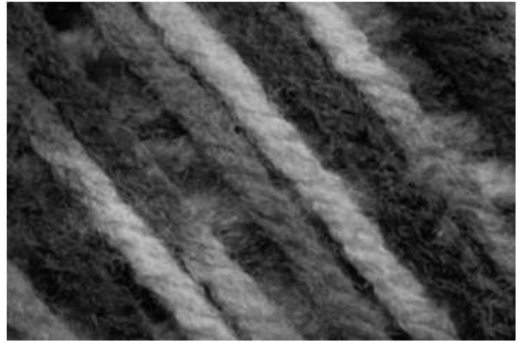

Gambar-4 Citra kain tenun dari Gambar-2 yang telah dilakukan operasi modifikasi histogram dengan menggunakan prosedur *averaging mask*.

*Median filtration* mengacu pada opsi untuk meratakan fungsi *brightness*, dan didasarkan untuk menentukan nilai piksel baru selain dengan melakukan prosedur *averaging mask*. *Median filtration* tidak didasarkan pada penggunaan nilai piksel rata-rata dari sekitar yang dipilih, namun dengan menerima nilai *brightness* terdekat yang ada di sekitarnya. Filtrasi median menekankan dan menandai kontur yang ada pada gambar. Contoh penerapan prosedur ini disajikan pada Gambar-5. Dengan menggunakan m*edian filtration* bersama-sama dengan *average filter*, kita dapat memperoleh efek yang lebih baik daripada tanpa filter ini, karena dapat meningkatkan efisiensi alokasi kontur. *Median filter* dapat digunakan pada semua mode warna, kecuali mode RGB 48-bit, kisaran abu-abu 16 bit, mode warna dengan palet, dan mode hitam-putih.

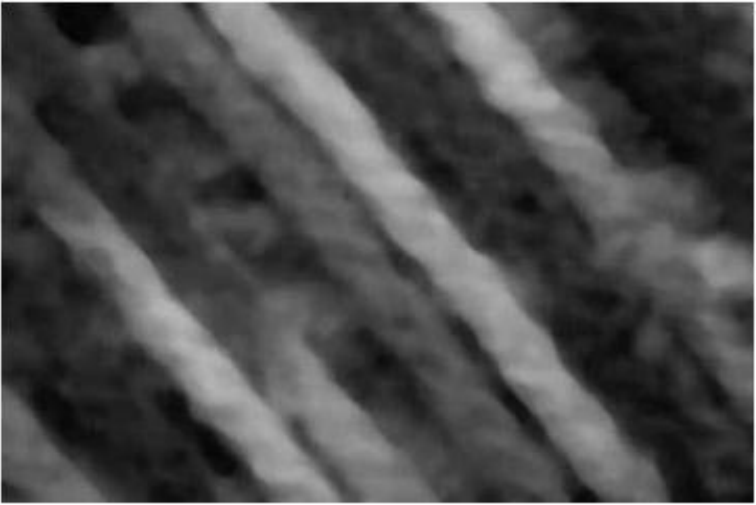

Gambar-5 Citra kain tenun dari Gambar-2 yang telah dilakukan operasi *median filtration*.

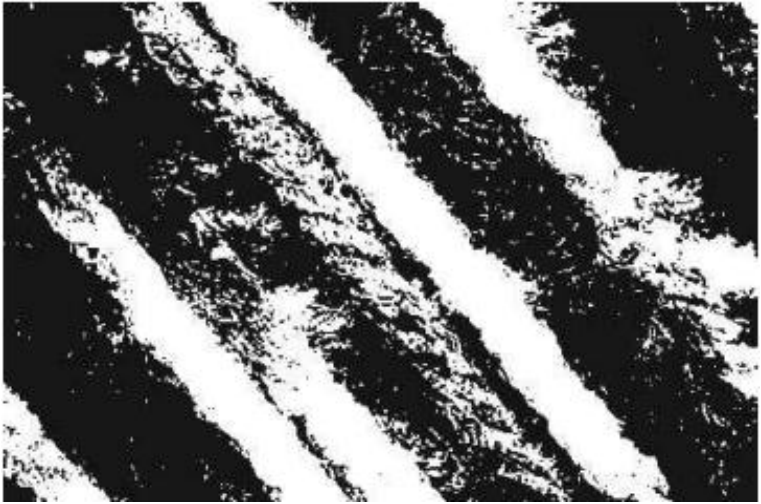

Gambar-6 Citra kain tenun dari Gambar-2 yang telah dilakukan prosedur *threshold*.

*Threshold* merupakan salah satu metode gradient yang digunakan untuk mengekstraksi kontur dari citra yang dianalisis. Hal ini didasarkan pada perubahan nilai fungsi *brightness* dari piksel citra digital tertentu. Banyak jenis *mask* yang bisa digunakan untuk menerapkan metode gradien. *Filter Laplace* adalah contoh filter gradien yang memungkinkan kontur diekstraksi, sementara pada saat bersamaan tetap mempertahankan *brightness* sebelumnya di dalam area yang ditandai. Proses *thresholding* pada skala abu-abu memungkinkan dilakukannya segmentasi citra. Hal ini dapat dilakukan dengan menerapkan proses di mana *gray level* dari permukaan gambar dianalisis dibandingkan dengan *gray level* yang ditentukan. Jika tingkat abu-abunya lebih tinggi dari ambang batas yang ditetapkan, maka setiap area diterima sebagai putih, jika tingkat abu-abu lebih kecil dari nilai ambang batas yang ditentukan, maka area akan diterima sebagai warna hitam. Prosedur *threshold* memungkinkan nilai batas *brightness* ditentukan dan ditetapkan, dengan kata lain, kita dapat menetapkan ambang batas untuk filtrasi yang akan kita gunakan. Piksel dengan nilai lebih tinggi atau lebih rendah dari nilai ambang diproyeksikan putih atau hitam tergantung pada pilihan yang dipilih. Sisa piksel tidak berubah, dan mempertahankan warna sebelumnya. Pilihan "*both level*" akan menyebabkan semua piksel berubah menjadi putih atau hitam, sesuai dengan nilai relasi *brightness* terhadap nilai ambang yang ditentukan. Ambang batas dapat ditentukan untuk keseluruhan mask pada citra digital yang dianalisis atau untuk *channel color* yang dipilih saja [2]. Contoh penerapan prosedur threshold disajikan pada Gambar-6.

Autokorelasi adalah teknik pengolahan citra yang berbeda. Teknik Autokorelasi menggabungkan semua fragmen gambar yang dianalisis, dan sering digunakan untuk mengkarakterisasi struktur *repeated mask* dari citra yang dianalisis. Penerapan autokorelasi menciptakan kemungkinan untuk menentukan dimensi rata-rata unit berulang dari *mask* yang dianalisis dari suatu objek. Teknik ini memudahkan untuk mereproduksi unit piksel berulang dalam kaitannya dengan keseluruhan gambar yang dianalisis [3]. Gambar yang diproses oleh prosedur ini disajikan pada Gambar-7.

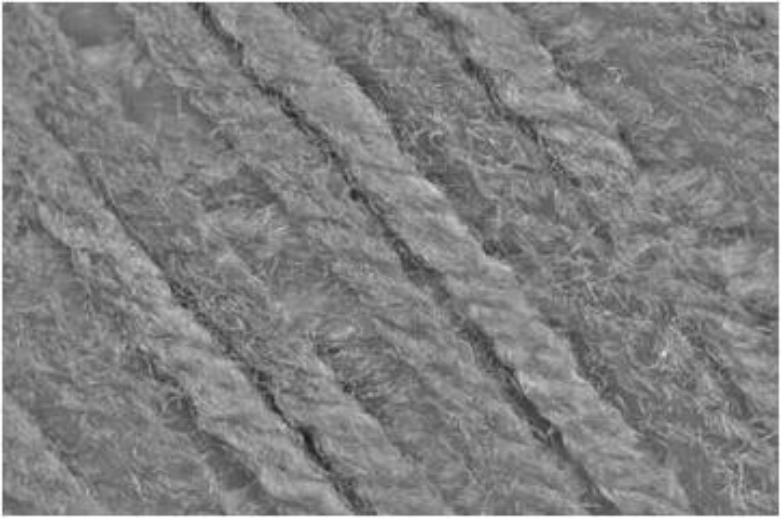

Gambar-7 Citra kain tenun dari Gambar-2 setelah penerapan teknik pengintegrasian *mask structure* dengan menggunakan prosedur *autocorellation.*

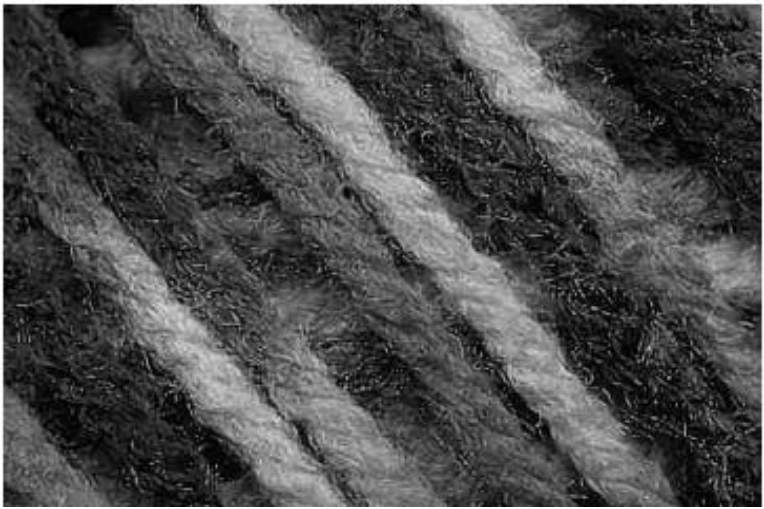

Gambar-8 Citra kain tenun dari Gambar-2 setelah penerapan teknik metode frekuensi barbasis *fourier transform.*

Metode frekuensi, teknik ini didasarkan pada modifikasi transformasi *Fourier* dari fungsi intensitas. Perbaikan citra yang dianalisis diperoleh dengan menentukan *reverse transform*. Metode frekuensi memerlukan daya perhitungan yang besar, karena penyaringan mencakup semua titik gambar di ruang frekuensi, dan tidak hanya beberapa piksel-piksel di bawahnya yang terbatas. Metode frekuensi memungkinkan faktor-faktor tersebut dapat dieliminasi atau dikompensasikan, seperti pencahayaan yang tidak homogeny dan kesalahan geometris dari jalur perolehan gambar. Penerapan *low filter* dan *high filter* memungkinkan fungsi intensitas dan kontur yang ditandai untuk diratakan. Contoh penerapan metode ini disajikan pada Gambar-8.

*Erosion* dan *dilatation* adalah salah satu operasi morfologi yang umum digunakan untuk memperbaiki citra yang dianalisis. Prosedur koreksi *erotion* dan *dilatation* didasarkan pada penambahan atau penghilangan piksel dari *mask* gambar *biner*, sesuai dengan peraturan yang diformulasikan berdasarkan standar yang diperoleh dari piksel tetangga. Contoh ditunjukkan pada Gambar 9.

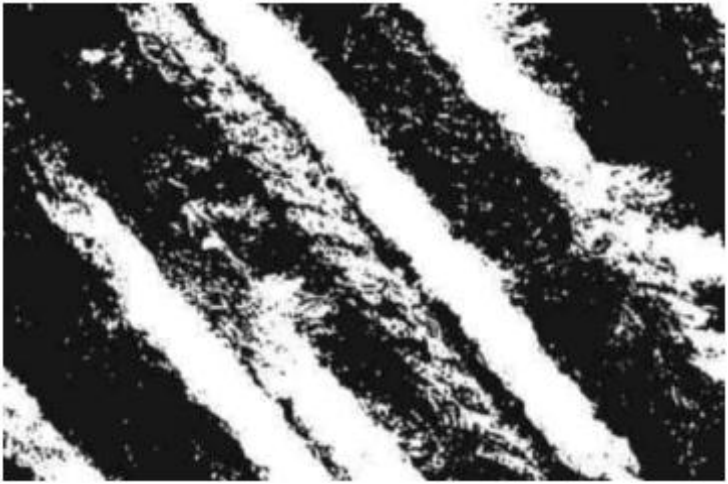

Gambar-7 Citra kain tenun dari Gambar-2 setelah penerapan prosedur *erotion* dan *dilatation* pada *mask* yang sebelumnya telah dilakukan proses *threshold.*

## 4. PEMULIHAN FITUR CITRA DIGITAL

Pemulihan (*reinstating*) [4] digunakan untuk menghilangkan dan meminimalkan fitur gambar yang menurunkan kualitasnya. Mendapatkan gambar dengan metode optik, opto-elektronik atau elektronik melibatkan degradasi beberapa fitur gambar yang tidak dapat dihindari selama proses pendeteksian. Selain itu, *noise internal* pada sensor gambar, pengaburan gambar yang disebabkan oleh kamera yang tidak fokus, serta turbulensi dan polusi udara di atmosfer sekitarnya dapat menyebabkan memburuknya kualitas [1]. Mengembalikan fitur gambar yang diinginkan berbeda dari perbaikan citra, yang prosedurnya terkait dengan penyorotan atau penyemaran fitur khas gambar yang ada. Mengembalikan fitur gambar yang diinginkan terutama mencakup koreksi berikut:
- mengembalikan ketajaman yang turun sebagai akibat dari fitur sensor yang kurang baik, atau lingkungan yang menyebabkan ketajaman gambar menurun
- *noise filtration*,
- *distortion correction*, dan
- koreksi *nonlinearity* sensor.

## 5. KOMPRESI DATA CITRA DIGITAL

Kompresi data citra digital [1] didasarkan pada meminimalkan jumlah byte yang diminta untuk merepresentasikan gambar. Efek kompresi dicapai dengan mengubah citra digital yang diberikan ke tabel angka yang berbeda sedemikian rupa sehingga jumlah informasi awal dimasukkan ke dalam jumlah sampel yang lebih sedikit.

## 6. PENERAPAN TEKNIK PENGOLAH CITRA UNTUK MENGANALISIS DAN MENGENALI OBJEK

Berdasarkan tinjauan literatur, dapat dinyatakan bahwa teknik pengolahan citra didasarkan pada penggunaan perangkat dan sistem optoelektronik analog dan digital yang memungkinkan gambar dengan distribusi informasi khusus ditempatkan pada input atau output sistem. Terlepas dari unit pengolahan gambar analog dan digital, teknik pengolahan gambar mencakup pengenalan gambar dan grafis komputer. Analisis dan pengenalan gambar dihubungkan dengan deteksi dan pemrosesan gambar, proyeksi, transmisi dan penyimpanan informasi, serta pengenalan dan pembangkitan citra [5]. Bidang teknik interdisipliner yang luas untuk teknik gambar mencakup lebih banyak lagi bidang pengetahuan teoritis dan eksperimental, teknologi, sistem, perangkat keras, dan perangkat lunak yang harus digunakan. Perkembangan teknik pengolahan citra yang dinamis berarti penerapan teknik tersebut diperluas, dan mencakup bidang sains lainnya, termasuk ilmu tekstil, yang dihasilkan dari pengembangan signifikan menjadi bidang seperti rekayasa material dan teknologi, serta teknik digital dan teknik komputer.

Pengolahan citra bukanlah bidang baru teknik pengolahan. Selama periode awal yang panjang, optik memiliki dampak mendasar pada pembentukan dan pengembangan bidang sains ini. Saat ini, informatika (ilmu pengetahuan dan teknologi informasi) dan teknologi mikroprosesor telah banyak mempengaruhi perkembangan yang pesat dalam dunia teknik pengolahan citra. Teknik pengolahan citra meliputi hal berikut:

- Pengolahan gambar, yang mencakup mengubah gambar, yaitu gambar dengan fitur tertentu, menjadi gambar dengan fitur lain yang diinginkan,
- Pengenalan citra, dengan tujuan untuk mengidentifikasi fitur dan objek gambar yang dipilih yang menjadi *subject of interest*; Pengenalan gambar memungkinkan pemilihan gambar, dan
- Grafik komputer dengan tujuan membuat gambar berdasarkan deskripsi yang diasumsikan.

## 7. TINJAUAN SINGKAT TENTANG TEKNIK PENGOLAHAN CITRA YANG DIRANCANG UNTUK MENGIDENTIFIKASI BAHAN TEKSTIL

Analisis digital dapat diterapkan untuk mengidentifikasi dan mengukur dimensi geometris benda tekstil dengan dimensi sangat kecil; khususnya memungkinkan struktur objek diselidiki untuk dianalisis. Proses identifikasi parameter struktural merupakan hal yang penting [6, 7]. Berdasarkan literatur yang terkait dengan masalah ini, kita dapat menyatakan bahwa teknik tersebut dapat digunakan untuk menentukan diameter serat. Selanjutnya, dengan menggunakan teknik tersebut memungkinkan untuk mendapatkan pandangan dari objek tekstil benang untuk mengamati kemungkinan cacat benang dan menentukan penyebabnya, terutama pandangan ujung sambungan benang, sehingga dapat diukur secara ilmiah dan kuantitatif mengenai hasil penyambungan benangnya. Dengan menggunakan teknik ini, peneliti dapat mengukur parameter struktural dasar produk tekstil benang, seperti *thickness*, *hairiness*, dan jumlah antihan [6]. Teknik yang dibahas juga memungkinkan dilakukannya pengamatan serupa pada objek tekstil 2D dan 3D.

Pohole [7] pertama kali menerapkan analisis citra komputer digital di bidang tekstil. Pada 1970-an, Pohole menggunakannya untuk memperkirakan daerah penampang serat wol. Cara baru menerapkan teknik pengolahan gambar menciptakan kemungkinan untuk menganalisis hasil pengukuran area penampang serat, memperkirakan penyimpangan campuran serat pada permukaan campuran benang, *maturity* serat, dan kerusakan pada serat wol. Masalah ini telah menjadi subyek penelitian yang dilakukan oleh Berlin, Worley, Raey [8], Thibodeaux, Evans [9], Watanabe, Kurosaki, Konoda [10], Zhao, Johnson, Willard [11], Żurek, Krucińska, dan Adrian [12]. Perkembangan teknik pengolahan citra yang pesat pada tahun 1990an membuka jalan bagi penelitian mengenai prosedur baru analisis ini.

Algoritma untuk digitalisasi gambar berfungsi untuk memperkirakan fitur morfologis *nonwoven* seperti porositas, distribusi orientasi serat dalam *nonwoven*, dan estimasi distribusi keteraturan serat di jaring telah dikembangkan oleh Huang dan Bresee [15]. Peneliti ini juga menerapkan proses pengukuran otomatis berdasarkan prosedur koreksi citra dengan menggunakan prosedur thresholding skala abu-abu.

Zhang dan Bresee [16] membandingkan berbagai teknik analisis citra yang bertujuan untuk mengenali dan mengklasifikasikan dua jenis cacat, yaitu sambungan dan *thick place*s yang terjadi pada kain tenun. Mereka menerapkan segmentasi citra menggunakan nilai ambang batas dari *mask* objek. Secara mandiri, mereka juga melakukan prosedur peningkatan kualitas gambar dengan menggunakan operasi koreksi seperti *histogram average*, *autocorelation*, *erotion*, dan *dilatation*. Mereka menyatakan bahwa penerapan koreksi citra dengan penggunaan metode penentuan secara statistik *grey level* gambar (prosedur *threshold*) lebih efisien daripada operasi morfologi yang menggunakan prosedur sederhana untuk menghilangkan perbedaan *mask* objek dengan *erotion* atau *dilatation*. Menurut Zhang dan Bresee [16], penerapan metode morfologi untuk teknik pengolahan citra memerlukan perhitungan daya yang lebih besar dibandingkan dengan menggunakan metode statistik, mengingat kualitas pemrosesan *mask* gambar yang diminta lebih tinggi.

Cybulska [6], [17] juga mengusulkan metode sendiri untuk memperkirakan struktur benang dengan menggunakan analisis citra digital. Dia melakukan penilaian terhadap parameter struktur dasar benang seperti ketebalan, rambut, dan twist, dengan menerapkan teknik pengolahan citra yang diperluas dengan metode numerik. Metode yang diusulkan memungkinkan karakteristik struktural numerik diperoleh pada setiap titik panjang benang, serta nilai rata-rata dan ukuran dispersi yang dapat diterima untuk parameter struktural benang.

Selanjutnya, Kopias, Mielicka dan Stempień [21] menggunakan digitalisasi citra untuk mengevaluasi poliuretan dan sambungan benang bertekstur. Untuk digitasi gambar mereka menerapkan metode berdasarkan pemindai yang terhubung ke stand komputer yang dilengkapi dengan program perangkat lunak yang dirancang untuk pengenalan objek otomatis. Kelainan pada proses pengenalan citra otomatis dieliminasi secara manual.

Teknik segmentasi, yang menggabungkan tingkat pengolahan citra awal dengan analisis *threshold* dari objek tertentu serta digunakan dalam analisis citra komputer. Teknik tersebut memungkinkan pemilihan *area* pada gambar yang memenuhi kriteria homogenitas yang ditentukan dari *mask*; Hal tersebut berarti akan membedakan objek dalam proses digitalisasi yang berbeda dari ambang *threshold* yang telah diatur. Analisis komparatif metode segmentasi yang dipilih berdasarkan gradien intensitas warna dilakukan oleh Krucińska dan Graczyk [22], yang mengukur luas permukaan serat. Berdasarkan penyelidikan Materka [23], mereka mengukur jumlah piksel yang termasuk pada objek yang diberikan, dan selanjutnya dikalikan dengan *area* nyata dalam piksel. Analisis menunjukkan bahwa segmentasi berdasarkan gradien warna intensitas piksel tetangga menghasilkan hasil yang hampir sama dengan segmentasi manual. Ini berasal dari prinsip kerja yang sama dari kedua algoritma dan akal manusia, karena manusia yang melihat objek membedakan batasnya terlebih dahulu di tempat-tempat yang ditandai oleh perubahan intensitas warna yang terbesar. Menurut Woźnicki [1], resolusi jalur elektronik dari digitalisasi gambar topeng memiliki pengaruh terbesar pada resolusi akhir citra digital dan proses digitalisasi, dan resolusi sistem optik hanya pada tingkat yang lebih rendah. Woźnicki, dalam bukunya mengenai dasar-dasar teknik pemrosesan citra [1], juga menjelaskan sumber kesalahan perolehan gambar, yang menunjukkan perlunya perhatian khusus diberikan pada kesalahan geometri gambar. Kesalahan ini sangat penting dalam pengolahan sampling; Mereka diwakili dalam deformasi jaring segi empat yang diterima sebagai gambar uji, dan mewakili penyimpangan distorsi. Deformasi gambar terbesar terjadi pada batas gambar, dan ini adalah alasan bahwa pengukuran harus dilakukan di tengah gambar. Iluminasi juga memiliki pengaruh besar terhadap kualitas perolehan citra digital. Menurut Kopias dan Jurasz [24, 25], saat mengukur geometri tekstil, sangat menguntungkan untuk menganalisis gambar struktur datar yang diperoleh dengan iluminasi dengan cahaya yang dipantulkan dan dipancarkan. Menggunakan sistem optik dengan transduser CCD memungkinkan hasil terbaik diperoleh dengan perolehan kembali gambar; deformasi hanya bisa terjadi pada margin *mask* gambar. Menurut Investigasi ke dalam geometri objek standar dan estimasi *inhomogeneity* mereka telah dilakukan oleh Perzyna, yang menjelaskan masalah ini di tesis Ph.D. miliknya [27].

## 8. KESIMPULAN

Perkembangan teknik komputer menawarkan banyak kesempatan untuk penerapannya dalam ilmu dan praktik tekstil. Penggunaan analisis citra komputer, telah memungkinkan identifikasi dimensi geometris dari benda tekstil sangat kecil. Menggunakan teknik koreksi gambar memungkinkan penghapusan kesalahan struktural pada kain yang sebelumnya akan diabaikan. Menerapkan teknik koreksi gambar memungkinkan identifikasi rinci struktur dan geometri dari kain tekstil benang. Menguraikan algoritma digitalisasi, dikombinasikan dengan metode numerik, memungkinkan diperolehnya karakteristik numerik dari struktur produk tekstil.

Berdasarkan metode analisis citra digital yang telah diterapkan sebelumnya dalam ilmu dan praktik tekstil, dan sebagai hasil dari banyak usaha, penulis telah mengembangkan metode analisis citra digital asli yang dirancang untuk memperkirakan parameter sambungan benang sebagai hasil dari proses penyambungan benang. Prosedur analisis citra digital yang digunakan memungkinkan estimasi parameter struktur eksternal sambungan benang dengan pengukuran pada panjang yang tak terbatas berserta dengan karakteristik numeriknya. Citra yang diperoleh sebagai hasil analisis citra komputer dapat digunakan untuk penyelidikan lebih lanjut mengenai estimasi kualitas sambungan benang, dan dapat memungkinkan dilakukan pengenalan secara cepat terhadap kemungkinan adanya kesalahan penyambungan benang tanpa simpul.

## 9. REFERENCE

[1] Woźnicki, J. 'Podstawy techniki przetwarzania obrazu', Wydawnictwa Komunikacji i Łączności, Warszawa 1996r.
[2] www.corel.com
[3] Zhang Y.F., Bresee, R. R. 'Fabrik Detection and Classification Using Image Analysis', Textile Research Journal, 65, 1, 1995, pp.1-9.
[4] Katsaggelos A. K. (ed.), Digital Image Restoration. Springer-Verlag, Berlin 1991.

[5] Górkiewicz-Gulwas, H., Woźnicki, J., 'Contemporary significance of the 'image technique' notion and its relation to other notions connected with them (in Polish)', Kwartalnik Elektroniki i Telekomunikacji, 41, 1995, pp.131.

[6] Cybulska, M., 'Analysis of Warp Destruction in the Process of Weaving Using the System for Assessment of the Yarn Structure', Fibres & Textiles in Eastern Europe, vol. 5, No 4., 1997, pp. 68 – 72.

[7] Pohle E., Interlaboratory Test for Wool Fineness Using the PiMc.', J. Testing Eval., 3, 1975, pp. 24-26.

[8] Berlin J., Worley S., Ramey H., 'Measuring the Cross-Sectional Area of Cotton Fibers with an Image Analyzer', Textile Research Journal, 51, 1981, pp. 109-113.

[9] Thibodeaux D., Evans J., 'Cotton Fiber Maturity by Image Analysis', Textile Research Journal, 56, 1986, pp. 130-139.

[10] Watanabe A., Kurosaki S., Konoda F. 'Analysis of Blend Irregularity in Yarns Using Image Processing', Textile Research Journal, 62, 1992, pp. 729-735.

[11] Zhao W., Johnson N., Willard A., 'Investigating Wool Fiber Damage by Image Analysis' Textile Research Journal, 56, 1986, pp. 464-466.

[12] Żurek W., Krucińka I., Adrian H. 'Distribution of Component Fibers on the Surface of Blend Yarns' Textile Research Journal, 52, 1982, pp. 473-478.

[13] Wood E. 'Applying Fourier and Associated Transforms to Pattern Characterization in Textiles', Textile Research Journal, 60, 1991, pp. 212-220.

[14] Wu Y., Pourdeyhimi B., Spivak M. 'Texture Evaluation of Carpets Using Image Analysis' , 61, 1991, pp. 407-419.

[15] Huang X., Bresee R., 'Characterizing Nonwoven Web Structure Using Image Analysis Techniques', INDA, 5, 1993, pp. 143 – 21.

[16] Zhang Y. F., Bresee R.R., 'Fabric Defect Detection and Classification Using Image Analysis', Textile Research Journal, 65, 1995, pp. 1 – 9.

[17] Cybulska M., 'Assessing Yarn Structure with Image Analysis Methods', Textile Research Journal, 69, 1999, pp.369-373.

[18] Masajtis J., Cybulska M., 'Computer estimation of the thread twist' (in Polish), Proceedings of the IMTEX' 95 International Scientific Conference, Technical University of Łódź, 22 – 23 May 1995.

[19] Masajtis J., 'Thread Image Processing in the Estimation of Repetition of Yarn Structure', Fibres & Textiles in Eastern Europe, vol. 10, No. 4, 1997, pp.68-72.

[20] Krucińska I., Kruciński S., 'Evaulating Fibrous Architecture of Nonwovens with ComputerAssigned Microscopy', Textile Research Journal, 69, 1999, pp. 363-369.

[21] Kopias K., Mielicka F., Stempień Z., 'An attempt to estimate spliced yarn using computer image analysis' (in Polish', IMTEX' 98 International Scientific Conference, Technical University of Łódź, June 1998.

[22] Krucińska I., Graczyk M., 'Review of selected methods of image analysis used for quality estimation of fibres and yarns (in Polish)', Proceedings of the 2nd Scientific Conference of the Textile faculty of the Technical University of Łódź, 1999, J-22, pp. 5 – 8.

[23] Materka A., 'Elementy cyfrowego przetwarzania i analizy obrazów' PWN, Warszawa 1991.

[24] Jurasz J., 'Estimation of the woven fabric structure during the weaving process (in Polish)', Przegląd Włókienniczy, No. 6, 1994, pp.235 – 237.

[25] Jurasz J., 'Application of the stereophotogrammatry method for identification of structure distortions of raw fabrics' (in Polish), Przegląd Włókienniczy, No. 7, 1994, pp.269 – 274.

[26] Tadeusiewicz R., Korohoda P., 'Komputerowa analiza i przetwarzanie obrazów', Wydawnictwo Fundacji Postępu Telekomunikacji , Kraków 1997. pp. 243-265.

[27] Perzyna M. Integrated estimation method of geometrical and mechanical properties of knitted fabrics by a system computer image analysis (in Polish). Ph.D. thesis, Technical University of Łódź, Branch in Bielsko-Biała, 2000.

[28] Bissman D., 'Knotfree – Yarn joins by splicing', Int. Textile Bull. Spinning, No. 3, 1981.

[29] Mashaly, R., Helw, E. El-Aki, K., 'Factors Affecting Spliced Yarn Quality', India. Textile J. Journal, 8, 1990, pp. 60-64.
[30] Cheng K.P.S., H..L.. Lam H.L., 'Physical Properties of Pneumatically Spliced Cotton Ring Spun Yarns' – Textile Research Journal, 70, 2000, pp. 1053-1057.
[31] Cheng K.P.S., H..L.. Lam H.L., 'Strength of Pneumatic Spliced Polyester/Cotton Ring Spun Yarns', Textile Research Journal, 70, 2000 pp. 243-246.
[32] Cheng K.P.S., Chan, K.K., How Y.L. and Lam H. L. I., 'Spliced Yarn Qualities—Breaking Strength and Flexural Rigidity of Spliced Yarns', Textile Asia 5, 1997, pp.45-47.
[33] Drobina R., Machnio M., 'Estimation of the properties of spliced combed woollen yarn-ends' (in Polish), Przegląd Włókienniczy i Technik Włókienniczy, vol. 54, No. 10, 2000, pp.3 –7.
[34] Drobina R., Machnio M., 'Visual estimation of spliced combed woollen yarn-ends' (in Polish), Przegląd Włókienniczy i Technik Włókienniczy, vol. 55, No. 1, 2001, pp. 12 – 14.
[35] Gebald G., 'Splicing technology in auto-winding', Textile Month, 7,1982.
[36] Kaushik R.C.D., Hari P.K., 'Performance of spliced yarn in warping and weaving', Textile Research Journal, 57, 1987, pp.670.
[37] Kuo Hung-Feng Jeffrey, Lee Ching-Jeng, 'A Back-Propagation Neural for Recognizing Fabric Defects', Textile Research Journal, 73, 2003, pp. 147-151.
[38] Chen, P. W., Liang, T. C., Yau, H. F., 'Classifying Textile Faults with a Back-Propagation Neural Network Using Power Spectra', Textile Research Journal 68, 1998, pp. 121-126.
[39] Cheng K.P.S., Lam H.L.L.., 'Evaluating and Comparing the Physical Properties of Spliced Yarns by Regression and Neural Network Techniques', Textile Research Journal 73, 2003, pp. 161-164.
[40] Huang Chang-Chiun, Chen I-Chun, 'Neural-Fuzzy Classification for Fabric Defects', Textile Research Journal 71, 2001, pp. 220-224.
[41] Chiu Shih-Hsuan, Chen Hung-Ning, Chen Jyh-Yeow., 'Appearance Analysis of False Twist Textured Yarn Packages Using Image Processing

and Neural Technology', Textile Research Journal, 71, 2001, pp. 313-317.
[42] Sakaguchi A., Guang Hua-Wen, Matsumoto Y, Toriumi K., Kim Hyungsup., 'Image Analysis of Woven Fabric Surface Irregularity', Textile Research Journal, 71, 2001, pp. 666-671
[43] Wen Che-Yen., 'Defect Segmentation of Texture Images with Wavelet Transform and a Cooccurrence Matrix', Textile Research Journal, 71, 2001, pp. 743-749.
[44] Kuo Chung-Feng Jeffrey, Lee Ching-Jeng, Tsai Cheng-Chih., 'Using a Neural Network to Identify Fabric Defects in Dynamic Cloth Inspection', Textile Research Journal, 73, 2003, pp. 238-244.
[45] Kuo Chung-Feng Jeffrey, Shih Chung-Yang, Kao Chih-Yuan, Lee Jiunn-Yih., 'Color and Pattern Analysis of Printed Fabric by an Unsupervised Clustering Method', Textile Research Journal, 75, 2005, pp. 9-12.
[46] Documentation of the Microscan 1.5 system for computer image analysis.
[47] Reports of research work financed from the resources for science provided for the years 2005 and 2006, as the promoter research project No. 3 T08E09629 directed by Mieczysław S. Machnio Ph.D., D.Sc., Professor of Bielsko-Biała University.